\documentclass[a4paper,conference]{IEEEtran}

\usepackage{cite}

\usepackage{tikz}
\usetikzlibrary{arrows,snakes,shapes,positioning,arrows,decorations.markings,calc}
\usetikzlibrary{dsp,chains}
\usepackage{psfrag}
\usepackage{color}
\definecolor{green}{rgb}{0,0.4,0.05}
\definecolor{red}{rgb}{0.8,0,0}

\usepackage{pgfplots}
\pgfplotsset{compat=newest}
\pgfplotsset{plot coordinates/math parser=false}
\newlength\figureheight
\newlength\figurewidth
\pgfplotsset{every axis plot/.append style={line width=0.8pt}}

\usepackage{amsmath}

\usepackage{authblk}
\usepackage{comment}
\usepackage{algorithm}
\usepackage{algorithmic}

\newcommand{\vect}[1]{\mathbf{#1}}
\newcommand{\matt}[1]{\mathbf{#1}}
\newcommand{\jim}{\mathrm{j}\,}

\DeclareMathOperator*{\argmax}{arg\,max}

\DeclareMathOperator*{\E}{E}
\newcommand{\T}{\operatorname{\mathrm{T}}}
\newcommand{\Q}{\operatorname{\mathcal{Q}}}
\newcommand{\He}{\text{H}}
\newcommand{\diag}{\text{diag}}

\begin{document}

\title{Spatial Coding Based on Minimum BER in 1-Bit Massive MIMO Systems}

\author[1]{Hela~Jedda}
\author[2]{Amine~Mezghani}
\author[1]{Jawad~Munir}
\author[1]{Fabian~Steiner}
\author[1,3]{Josef~A.~Nossek}
\affil[1]{Technical University of Munich, 80290 Munich, Germany}
\affil[2]{University of California, Irvine, Irvine, CA 92697, USA}
\affil[3]{Federal University of Cear\'a, Fortaleza, Brazil}
\affil[ ] {Email: hela.jedda@tum.de, amezghan@uci.edu, josef.a.nossek@tum.de}

% make the title area
\maketitle
\tikzset{DSP lines/.style={help lines,very thick,color=black}}
\tikzset{line_arrow/.style={help lines,very thick,color=black,->,-angle 90}}
\tikzset{filter/.style={rectangle,inner sep=0pt,minimum height=1cm,minimum width=1cm,draw=black,very thick}}
\tikzset{delay/.style={rectangle,inner sep=0pt,minimum size=1cm,draw=black,very thick}}
\tikzset{downsampling/.style={rectangle,inner sep=0pt,minimum height=0.8cm,minimum width=0.7cm,draw=black,very thick}}
\tikzset{upsampling/.style={rectangle,inner sep=0pt,minimum height=0.8cm,minimum width=0.7cm,draw=black,very thick}}
\tikzset{empty_node/.style={inner sep=0pt,minimum size=0cm}}
\tikzset{connection/.style={circle,draw=black,fill=black,inner sep=0pt,minimum size=2mm}}
\tikzset{coefficient/.style={isosceles triangle,draw=black,very thick,inner sep=0pt,minimum size=.7cm}}
\tikzset{source/.style={semicircle,minimum size=.5cm,draw=black,very thick,shape border rotate=270}}
\tikzset{adder/.style={circle,minimum size=.25cm,inner sep=0pt,draw=black,very thick}}
\tikzset{multiplier/.style={circle,minimum size=.25cm,inner sep=0pt,draw=black,very thick}}
\tikzset{double_arrow/.style={double distance=5pt,thick,shorten >= 6pt,decoration={markings,mark=at position 1 with {\arrow[scale=.6,>=angle 90]{>}}},postaction={decorate}}}
% As a general rule, do not put math, special symbols or citations
% in the abstract

\let\thefootnote\relax\footnotetext{This work was supported by German Academic Exchange Service (DAAD) with a DAAD scholarship for doctoral candidates (DAAD-Doktorandenstipendium).}
\begin{abstract}

We consider a downlink 1-bit quantized multi-user (MU) multiple-input-multiple-output (MIMO) system, where 1-bit digital-to-analog (DACs) and analog-to-digital converters (ADCs) are used at the transmitter and the receiver for economical and computational efficiency. We end up with a discrete memoryless channel with input and output vectors belonging to the QPSK constellation. In the context of massive (MIMO) systems the number of base station (BS) antennas is much larger than the number of receive antennas. This leads to high input cardinality of the channel. In this work we introduce a method to reduce the input set based on the mimimum bit-error-ratio (BER) criterion combined with a non-linear precoding technique. This method is denoted as spatial coding. Simulations show that this spatial coding improves the BER behavior significantly removing the error floor due to coarse quantization. 
\end{abstract}

% no keywords

% For peer review papers, you can put extra information on the cover
% page as needed:
% \ifCLASSOPTIONpeerreview
% \begin{center} \bfseries EDICS Category: 3-BBND \end{center}
% \fi
%
% For peerreview papers, this IEEEtran command inserts a page break and
% creates the second title. It will be ignored for other modes.
\IEEEpeerreviewmaketitle

\section{Introduction}
The 5G vision aims at having a fully connected network society, where the data is accessible everywhere and everytime  for everyone and everything. Data rates up to 10Gbps are envisioned and more than 50 billions of connected devices are expected \cite{5G}. Mobile communication will not be restricted to people anymore but will include all kinds of communication between connected entities, e.g. connected machines, vehicular telematics, automatic train control systems, industrial automation, e-health services. Some future applications like traffic safety and remote surgery require low latency times of less than one millisecond and high reliability. In addition, this technology has to be achieved with more energy efficient systems to have long-term sustainable technology. 

To meet the tremendous demand of higher data rates and traffic, two main research ideas are considered: increasing the number of antennas at the base station (BS), i.e. master node, that serve a smaller number of terminals, i.e. slave nodes, denoted by massive multiple-input-multiple-output (MIMO) \cite{LarssonMarzetta2014}, and using mm-Wave frequencies where higher bandwidth is still available \cite{mmWave5G}. 
Both directions lead to higher hardware energy consumption whether due to the large antenna arrays and thus the hardware needed for each antenna or due to the higher frequencies that the hardware have to operate at. 
Thus, the energy consumption becomes a crucial concern and has to be efficiently used to achieve reliable communication systems.
An important measure to achieve more energy efficient systems is the usage of 1-bit digital-to-analog converters (DACs) and analog-to-digital converters (ADCs). 

To meet the demand of low response times of future communication systems the coding task has to be less computationally complex and less time consuming, i.e. LDPC codes with small code lengths and less number of iterations for decoding. This comes, however, at the cost of the reliability. To solve this problem,  we introduce a channel spatial coding technique that ensures high reliability by mitigating the MU interference (MUI) and the quantization distortions.

The proposed spatial coding is different from the idea of spatial modulation (SM) \cite{SM2001, SMSiemens2012}. The SM is based on antenna multiplexing to transmit the information bits. The transmit antenna index is an information-bearing unit in addition to the symbol drawn from the constellation diagram. However, in our proposed technique all the transmit antennas are active and the set of transmit signals is reduced to a subset of signals that can be at best transmitted through the channel.

This paper is organized as follows: in Section \ref{sec:sysmodel} we present the system model. In Section \ref{sec:sc} we introduce the spatial coding method. In Sections \ref{sec:simresults} and \ref{sec:conclusion} we show the simulation results and summarize this work.

\textbf{Notation}: Bold letters indicate vectors and matrices, non-bold letters express scalars. The operators $(.)^{*}$, $(.)^{\rm T}$ and $(.)^{\rm H}$ stand for complex conjugation, the transposition and Hermitian transposition, respectively. The $n \times n$ identity (zeros, ones) matrix is denoted by $\mathbf{I}_{n}$ ($\mathbf{0}_{n},~ \mathbf{1}_{n}$). $\diag(\matt{A})$, $\det(\matt{A})$ denote a diagonal matrix containing only the diagonal
elements of $\matt{A}$ and the determinant of $\matt{A}$, respectively. The operator $\otimes$ denotes the Kronecker product.

\section{System Model}
\label{sec:sysmodel}

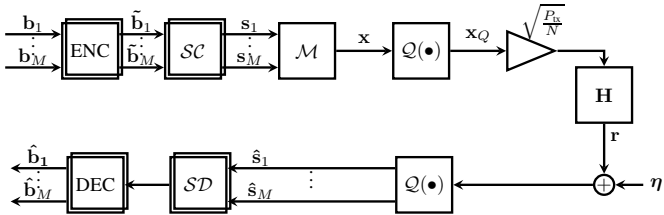
\begin{figure}[h!]
	\centering
	\resizebox{9cm}{!} {
\begin{tikzpicture}

\node (in){};
\node[filter] (encoding) [right=of in][xshift=0cm ] {ENC};
\node[filter] [right=of in][xshift=0.07cm, yshift=0.07cm] {};
\node[filter] (scoding) [right=of encoding] [xshift=-0.2cm ] {$\mathcal{SC}$};
\node[filter]  [right=of encoding][xshift=-0.13cm, yshift=0.07cm] {};
\node[filter] (mapping) [right=of scoding] {$\mathcal{M}$};
%\node[filter] (lut) [below=of mapping][yshift=0.5cm] {LUT};
\node[filter] (quantizer1) [right=of mapping] {$\mathcal{Q} (\bullet) $};
\node [coefficient](power)[right=of quantizer1]{};
\node [above=of power][xshift=0.4cm,yshift=-1.2cm] {$\sqrt{\frac{P_{\text{tx}}}{N}}$};
\node[filter] (channel) [below=of power] [xshift=1.5cm ][yshift=1cm ] {$\matt H$};
\node[adder] (noise) [below=of channel] [yshift=0.1cm ]  {$+$};
\node (n)[right=of noise][xshift=-0.5cm] {$\boldsymbol{\eta}$};
\node[filter] (quantizer2) [left=of noise][xshift=-1.5cm]  {$\mathcal{Q} (\bullet) $};
\node[filter] (sdecoding) [left=of quantizer2][xshift=-2cm ] {$\mathcal{SD}$};
\node[filter] [left=of quantizer2][xshift=-1.93cm, yshift=0.07cm  ] {};
\node[filter] (decoding) [left=of sdecoding][xshift=0.2cm] {DEC};
\node[filter]  [left=of sdecoding][xshift=0.27cm, yshift=0.07cm] {};
\node (out) [left=of decoding][xshift=-0.0cm ]{};

\draw[DSP lines] [-stealth] ([yshift=0.3cm]in.east) -- ([yshift=0.3cm]encoding.west)node[pos=0.5,above,yshift=-0.1cm]{$\vect b_1$}node[pos=0.5,below,yshift=0.3cm] {$\vdots$};
\draw[DSP lines] [-stealth] ([yshift=-0.3cm]in.east) -- ([yshift=-0.3cm]encoding.west)node[pos=0.5,above,yshift=-0.1cm]{$\vect b_M$};
\draw[DSP lines] [-stealth] ([yshift=0.3cm]encoding.east) -- ([yshift=0.3cm]scoding.west)node[pos=0.5,above,yshift=-0.1cm]{$\vect {\tilde b}_1$}node[pos=0.5,below,yshift=0.3cm] {$\vdots$};
\draw[DSP lines] [-stealth] ([yshift=-0.3cm]encoding.east) -- ([yshift=-0.3cm]scoding.west)node[pos=0.5,above,yshift=-0.1cm]{$\vect {\tilde b}_M$};
\draw[DSP lines] [-stealth] ([yshift=0.3cm]scoding.east) -- ([yshift=0.3cm]mapping.west)node[pos=0.5,above,yshift=-0.1cm]{$\vect s_1$}node[pos=0.5,below,yshift=0.3cm] {$\vdots$};
\draw[DSP lines] [-stealth] ([yshift=-0.3cm]scoding.east) -- ([yshift=-0.3cm]mapping.west)node[pos=0.5,above,yshift=-0.1cm]{$\vect s_M$};
\draw[DSP lines] [-stealth] (mapping.east) -- (quantizer1.west)  node[pos=0.5,above]{$\vect x$};
%\draw[DSP lines] [dashed] (lut.north) -- (mapping.south);
\draw[DSP lines] [-stealth] (quantizer1.east) -- (power.west)node[pos=0.5,above]{$\vect x_Q$};
\draw[DSP lines] [-stealth] (power.east) -| (channel.north);
\draw[DSP lines] [-stealth] (channel.south) -- (noise.north)node[pos=0.5,above,xshift=0.2cm]{$\vect r$};
\draw[DSP lines] [-stealth] (n) -- (noise.east);
\draw[DSP lines] [-stealth] (noise.west) -- (quantizer2.east);
\draw[DSP lines] [-stealth] ([yshift=0.3cm]quantizer2.west) -- ([yshift=0.3cm]sdecoding.east) node[pos=0.5,above,xshift=-0.9cm,yshift=-0.1cm]{$\vect {\hat s}_1$}node[pos=0.5,below,yshift=0.3cm] {$\vdots$};;
\draw[DSP lines] [-stealth] ([yshift=-0.3cm]quantizer2.west) -- ([yshift=-0.3cm]sdecoding.east) node[pos=0.5,above,xshift=-0.9cm,yshift=-0.1cm]{$\vect {\hat s}_M$};
\draw[DSP lines] [-stealth] (sdecoding.west) -- (decoding.east);
\draw[DSP lines] [-stealth] ([yshift=0.3cm]decoding.west) -- ([yshift=0.3cm]out.east)node[pos=0.5,above,yshift=-0.1cm]{$\vect{ \hat b_1}$} node[pos=0.5,below,yshift=0.3cm] {$\vdots$};
\draw[DSP lines] [-stealth] ([yshift=-0.3cm]decoding.west) -- ([yshift=-0.3cm]out.east)node[pos=0.5,above,yshift=-0.1cm]{$\vect{ \hat b}_M$} ;
\end{tikzpicture}%
}
	\caption{Coded 1-bit MIMO system model with QPSK symbols.}
	\label{fig:sysmodel}
\end{figure}

We consider a 1-bit downlink massive MU-MIMO scenario as depicted in Fig. \ref{fig:sysmodel} with $N$ transmit antennas at the BS and $M$ users each with $K$ antennas, where $N>>MK$. $M$ independent bit streams intended for each user are encoded separately with LDPC codes and then spatially coded to get the modulated signal vector $\vect s = \begin{bmatrix}
\vect s_1^{\T}& \vect s_2^{\T}& \cdots &\vect s_M^{\T}
\end{bmatrix}^{\T}$. The total coding rate of each user is denoted by $r_m = r_{m_{\text{LDPC}}} \cdot r_{m_{\text{SC}}}$, where $r_{m_{\text{LDPC}}}$ and $r_{m_{\text{SC}}}$ denote the LDPC coding rate and the spatial coding rate, respectively. The spatial coding will be explained in Section \ref{sec:sc}. The signal vector $\vect{s}_m = \begin{bmatrix}
s_{m_1}&  s_{m_2}& \cdots & s_{m_K}
\end{bmatrix}^{\T} \in \mathcal{O}^{K},~m=1,~\cdots,~M$, contains $K$ data symbols for the $m$-th user, where $\mathcal{O}$ represents the set of QPSK constellation. We assume that the entries of $\vect{s}$ are independent identically distributed (i.i.d.)
with zero mean and covariance matrix $\matt C_s = \sigma_{s}^2\matt{I}_{MK}$. 
In this system we deploy 1-bit quantization $\mathcal{Q}$ at the transmitter as well as at the receiver. The use of the 1-bit quantizer at the transmitter delivers a signal $\vect{x}_{Q} \in \mathcal{O}^N$. To mitigate the MUI and the distortions due to the coarse quantization, the input signal vector $\vect s$ is mapped to the unquantized transmit signal vector $\vect x$ prior to DAC. This mapping is based on a LUT of size $N\times 4^{MK}$, that is generated at the beginning of each coherence slot. 
The quantized signal $\vect{x}_{Q}$ gets scaled with $\sqrt{\frac{P_{\text{tx}}}{ N}}$, where $P_{\text{tx}}$ is the available power at the transmitter.
The received decoded signal vector $\hat{\vect{s}} = \begin{bmatrix}
\hat{\vect s}_1^{\T}& \hat{\vect s}_2^{\T}& \cdots &\hat{\vect s}_M^{\T}
\end{bmatrix}^{\T} \in \mathcal O^{MK}$ reads as $\hat{\vect{s}} = \Q \left(\sqrt{\frac{P_{\text{tx}}}{N}} \matt H \vect{x}_{Q} + \boldsymbol{\eta}\right)$, where $\matt H \in \mathrm{C}^{MK\times N}$ is the channel matrix
and $\vect{\boldsymbol{\eta}}\sim \mathcal{C}\mathcal{N}\left ( \vect{0}_{MK},\matt{C}_{\boldsymbol{\eta}} = \matt{I}_{MK} \right )$ is the AWG noise vector. We assume that the users' channels are uncorrelated but the $K$ antennas of each user are correlated with the correlation factor $\rho$. We get
\begin{align}
\E\{\matt H \matt H^{\He}\}= \left( (1-\rho )\matt I_K + \rho \mathbf{1}_K \right)  \otimes \matt I_M.
\end{align}

%\section{Mapping: $\mathcal{M}$}
%\label{sec:mapping}
%In this work, we do not design a precoder but we design the transmit vector signal $\vect x$ for a given input signal vector $\vect s$ depending on the channel, while we assume full CSIT. As depicted in Fig. \ref{fig:processing_steps}, first, an optimization problem is solved for all possible input vectors $\vect s$ to find the optimal transmit vectors $\vect x$. The used optimization problems are introduced in Section \ref{sec:optproblem}. Second, the solutions are stored in the LUT of size $N \times 4^M$. Since we are restricted to QPSK modulation, we get $4^M$ possible input vectors. Third, we map the given input vector $\vect s$ into a signal vector $\vect x$ according to the LUT, which is updated for each channel. The LUT approach is useful in case of small or moderate $M$ due to the small required 1-bit memory for each signal components and since the optimization has not to be performed for every input.
%
%The aim of the optimization problem is to jointly minimize the IUI and the quantization distortions. The optimization criterion is the minimun BER (MBER) under the constraint that $\vect x \in \mathcal{O}^N$. This constraint leads to a linear behavior of the quantizer at the transmitter, e.g. $\vect x_Q = \vect x$. Thus, the quantization distortions at the transmitter are omitted.
\section{Spatial Coding}
\label{sec:sc}
\begin{figure}[h]
\centering  

\psfrag{Nx4M}[][]{$N\times L$}
\psfrag{NxK}[][]{$N\times L'$}
\psfrag{s}[][]{$\vect s$}
\psfrag{x}[][]{$\vect x$}
\psfrag{CF}[][]{$\Phi$}
\psfrag{b}[][]{$\vect{\tilde b}$}
\psfrag{4M}[][]{\tiny $ L$}
\psfrag{L'}[][]{\tiny $ L'$}
\psfrag{...}[][]{\tiny $\cdots$}
\psfrag{1}[][]{\tiny 1}
\psfrag{2}[][]{\tiny 2}
\psfrag{1.}[][]{ 1.}
\psfrag{2.}[][]{ 2.}
\psfrag{3.}[][]{ 3.}

\includegraphics[width=0.48\textwidth]{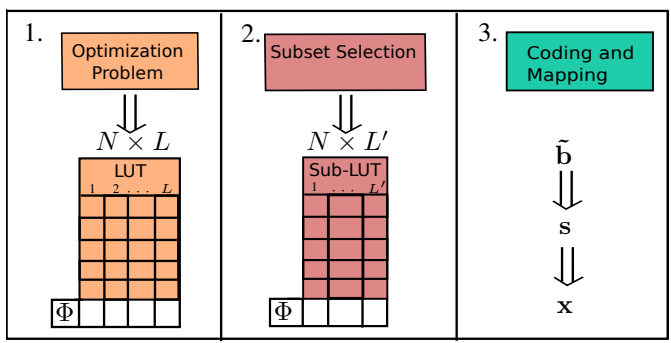}
\caption{Processing steps for each channel}
\label{fig:processing_steps}
\end{figure}

\subsection{Optimization Problem and Precoding}
This optimization problem was first inroduced in \cite{JeddaSAM2016} for precoding. The optimal transmit vector $\vect x$ is to be designed to achieve two goals: minimize the MUI and mitigate the quantization distortions. The problem formulation is given by
\begin{align}
\vect x = \argmax_{\tilde {\vect x} \in \mathcal O'^{N}} \Phi\left( \matt H, \tilde {\vect x}, \vect s\right),
\label{eq:opt_problem}
\end{align}
where 
\begin{align}
\Phi(\matt H, \vect x, \vect s) = \det \left( \Re \left\lbrace  \diag\left(\matt H \vect{x} \vect{s}^{\He} \right)^2 \right\rbrace \right).
\end{align}
We aim at getting the entries of the noiseless received signal $\vect r = \matt H \vect x$ in the same quadrants as the entries of the desired signal $\vect s$ and as far as possible from the quantization thresholds. We denote this design criterion as the minimum BER (MBER). The ideal constraint is $\vect x \in \mathcal{O}^N$, so that we get $\vect x_Q = \vect x$ and the quantization distortions are totally omitted. However, this leads to a non-convex solution set. The relaxed convex constraint $\vect x \in \mathcal{O}'^N$ makes sure that the elements of $\vect x$ belong to the box built by the QPSK constellation points $\mathcal{O}'$ and therefore the quantization distortions are minimized.

\begin{figure}[h]
\centering

\psfrag{M}[][]{$\mathcal{M}$}
\psfrag{S}[][]{$\mathcal{S}$}
\psfrag{X}[][]{$\mathcal{X}$}
\psfrag{s1}[][]{$\vect s^{(1)}$}  
\psfrag{x1}[][]{$\vect x^{(1)}$} 
\psfrag{s3}[][]{$\vect s^{(L)}$} 
\psfrag{x3}[][]{$\vect x^{(L)}$} 
\psfrag{...}[][]{$\cdots$}  
\includegraphics[width=0.4\textwidth]{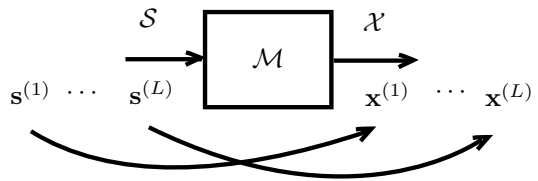}
\caption{Illustration of the mapping step.}
\label{fig:mapping}
\end{figure}

As shown in (\ref{eq:opt_problem}), the cost function to be maximized includes the input signal $\vect s$. Hence, the optimal transmit signal $\vect x$ that is obtained by solving (\ref{eq:opt_problem}) is the solution for one specific $\vect s$ from the alphabet $\mathcal{S}$. Since we deal with signals with QPSK entries, there are $4^{MK}$ possible distinct input vectors, i.e. $|\mathcal{S}| = L = 4^{MK}$. Therefore, the optimization problem has to be run for all $ L$ distinct input vectors $\vect s$ to get all $ L$ optimal transmit vectors $\vect x \in \mathcal{X}$. For illustration refer to Fig. \ref{fig:mapping}.  The optimal vectors $\vect x^{(1)},~ \vect x^{(2)},~\cdots,~\vect x^{(L)}$ are stored in the columns of a look-up table (LUT) of size $N\times L$ as shown in the first step of Fig. \ref{fig:processing_steps}. The resulting optimal cost function $\Phi$ is assigned to each optimal transmit vector $\vect x$ and thus to each column of the LUT. The LUT is updated for each coherence channel.

%
%\begin{figure}[h]
%\centering  
%\psfrag{Re}[][]{$\Re$}
%\psfrag{Im}[][]{$\jim \Im$}
%\psfrag{s}[][]{$\!\!s_m$}
%\psfrag{shat}[][]{$r_m$}
%\psfrag{d}[][]{$\:\delta_m$}
%\psfrag{phi}[][]{$\phi_m$}
%\psfrag{A}[][]{\textcolor{green}{$A_m = \Re\{r_m s_m^*\}$}}
%\psfrag{B}[][]{\textcolor{red}{$B_m=\Im\{r_m s_m^*\}$}}
%\includegraphics[width=0.25\textwidth]{figures/Visualization_delta}
%\caption{Illustration of the optimization problem}
%\label{fig:problem_illustration}
%\end{figure}
%
%\begin{align*}
%\sin\left(\frac{\pi}{4}-\vert\phi_m\vert\right)&=\frac{\delta_m}{\sqrt{A_m^2 + B_m^2}}, \phi_m \in \left]-\frac{\pi}{4}, \frac{\pi}{4}\right[ \\
%&= \frac{1}{\sqrt{2}}\left( \cos(\vert\phi_m\vert) - \sin(\vert\phi_m\vert) \right)\\
%\sin(\vert\phi_m\vert)&=\frac{\vert B_m \vert}{\sqrt{A_m^2 + B_m^2}}\\
%\cos(\vert\phi_m\vert)&=\cos(\phi_m)=\frac{ A_m }{\sqrt{A_m^2 + B_m^2}}
%\end{align*}
%
%\begin{align}
%\delta_m = \frac{1}{\sqrt{2}}  \left( A_m - \big \vert B_m \vert\right) 
%\end{align}
%
%$\delta_m$ is the minimal distance of the $m$th entry of the received signal to the quantization thresholds. 
\subsection{Subset Selection}
In this step we select a subset $\mathcal{S'}$ of $L'$ input vectors $\vect s$ and accordingly transmit vectors $\vect x$ that are best transmitted through the channel, i.e. that lead to the best cost functions $\Phi$.
\subsubsection{Single User}
The subset $\mathcal{S'}$ is selected as follows for the single user case
\begin{align}
\mathcal{S'} = \argmax_{\mathcal{S'} \subseteq \mathcal{S}} \sum_{\vect s \in \mathcal{S'}} \Phi(\matt H, \vect x, \vect s).
\end{align}
In the precoding task we store the optimal transmit vectors $\vect x$ for all input vectors in the LUT. Each column of the LUT leads to a certain cost function that can quantify the BER. When this LUT is sorted in descending order of the cost functions we can easily select the elements of the subset $\mathcal{S'}$ as the first $L'$ elements of the sorted LUT and get the sub-LUT of size $N \times L'$. This implies that we shape the input distribution probabilistically and assign the undesired input vectors with a probability value of 0 and the desired input vectors with equal probabilities as follows 
\begin{align}
P(\vect s) = \begin{cases} 1/L' &\text{ if } \vect s \in \mathcal{S'}\\ 0 &\text{  otherwise.}
\end{cases}
\end{align}
\subsubsection{Multi User}
For the multi user case the input vectors for each user have to be selected independently from the other users. The set of all possible input vectors is defined as
\begin{align}
\mathcal{S} = \prod_{m=1}^M \mathcal{S}_m,
\end{align}
where $\mathcal{S}_m$ represents the set of the possible input vectors for the $m$-th user.
The input alphabet cardinality for each user has to be reduced to $|\mathcal{S}_m'| = L_m'$, where $\mathcal{S'} =\prod_{m=1}^M \mathcal{S}_m'$ and $\vert \mathcal{S'}\vert = L' = \prod_{m=1}^M L_m'$. The optimal formulation to find $\mathcal{S}'$ is given by
\begin{align}
\mathcal{S'} = \argmax_{\mathcal{S'} \subseteq \mathcal{S}} \sum_{\vect s \in \mathcal{S'}} \Phi(\matt H, \vect x, \vect s) ~ \text{ s.t. } \mathcal{S'} =\prod_{m=1}^M \mathcal{S}_m'.
\label{eq:ss_mu}
\end{align}
However, solving (\ref{eq:ss_mu}) is not straight forward since it consists in a combinatorial optimization. To this end, we resort to a successive selection, which is sub-optimal but easy to solve. The selection steps are summarized in Algorithm \ref{algorithm:ss_mu}.
\begin{algorithm}
\caption{Find optimal $\mathcal{S'}$ for the multi user case}
\label{algorithm:ss_mu}
\begin{algorithmic}
\REQUIRE LUT, $L_m',~m=1,~\cdots,~M$
\ENSURE Sub-LUT, $S_m',~m=1,~\cdots,~M$
\STATE $\mathcal{S'} = \mathcal{S}$
\FOR {$m=1$ \TO $M$}
	\STATE $\mathcal{S}_m' = \argmax\limits_{\mathcal{S}_m' \subseteq \mathcal{S}_m} \sum\limits_{\vect s \in \mathcal{\tilde S}} {\rm E}_{\vect{s}_j \in \mathcal{S'}, j \neq m} \lbrace\Phi(\matt H, \vect  x, \vect { s})\rbrace$
	\STATE $\mathcal{S'} = \prod\limits_{i=1}^m \mathcal{S}_i' \times \prod\limits_{j=m+1}^M \mathcal{S}_j $
\ENDFOR
\end{algorithmic}
\end{algorithm}
Thus, the input distribution of the input vectors in each set $\mathcal{S'}$ is probabilistically shaped as follows
\begin{align*}
P(\vect s_m) = \begin{cases} 1/L'_m &\text{ if } \vect s_m \in \mathcal{S'}_m\\ 0 &\text{  otherwise.}
\end{cases}
\end{align*}
The spatial coding rate for each user is then defined as
\begin{align}
r_{m_\text{SC}} = \frac{\log_2(L_m')}{\log_2(L_m)}.
\end{align}
\subsection{Coding and Mapping}
\begin{figure}[h]
\centering  
\psfrag{Re}[][]{$\Re$}
\psfrag{Im}[][]{$\jim \Im$}
\psfrag{0}[][]{0}
\psfrag{1}[][]{1}
\psfrag{2}[][]{2}
\psfrag{3}[][]{3}
\includegraphics[width=0.2\textwidth]{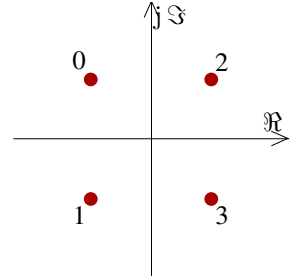}
\caption{Decimal representation of the QPSK symbols with Gray coding: $D_G(\bullet)$.}
\label{fig:qpsk_gray}
\end{figure}

After choosing the $L'$ input vectors to be transmitted through the channel, and hence choosing the $L_m'$ input vectors for each user, each encoded bit stream $\tilde{\vect b}_m$ has to be mapped to the possible input vectors $\vect s_m \in \mathcal{S}_m'$, $m=1,~\cdots,~M$. Each bit stream $\tilde{\vect b}_m$ is divided then into blocks of length $\log_2(L_m')$ and afterwards mapped to the input vectors according to an encoding scheme. At the receiver side, this encoding scheme has to be known so that the received signals can be decoded.  
%\begin{figure}[h]
%\centering  
%\psfrag{K}[][]{$L'$}
%\psfrag{log2(K)}[][]{$\log_2(L')$}
%\psfrag{Spatial Coding}[][]{Spatial Coding}
%\psfrag{bit stream}[][]{bit stream}
%\includegraphics[width=0.4\textwidth]{figures/spatial_coding}
%\caption{Spatial coding}
%\label{fig:spatial_coding}
%\end{figure}
We define the decimal value of the input vector $\vect s_m$ as
\begin{align}
\Vert \vect s_m \Vert_{10}= \sum_{k=1}^{K} 4^{k-1}  D_G(s_{m_k}),
\label{eq:decimal_value}
\end{align}
where $D_G(\bullet)$ gives the decimal representation of the QPSK symbol based on Gray coding as depicted in Fig. \ref{fig:qpsk_gray}. $D_G(\bullet)^{-1}$ is the inverse function and gives the corresponding QPSK symbol for a given decimal value from 0 to 3.
The $L_m'$ selected input vectors for each user are sorted in ascending order of the decimal value as defined in (\ref{eq:decimal_value}).
Each input vector $\vect s_m \in \mathcal{S}_m'$ is then encoded with the binary word of its position after the sorting. The first position is 0.
The mapping from the total input vector $\vect s$ to the transmit vector $\vect x$ is then performed based on the sub-LUT.
%The coding of the selected vectors is performed such that the code words of two vectors that have small hamming distance differ in a small number of bits. The vectors are sorted such that the hamming distance between two consecutive vectors is minimized.

\subsection{Example}
We assume that $M=2$ and $K=2$. Each user has $\vert \mathcal{S}_1 \vert  = L_1 = \vert \mathcal{S}_2 \vert  = L_2 = 16$ possible input vectors. The set $\mathcal{S} = \mathcal{S}_1 \times \mathcal{S}_2$ has a cardinality of $L=4^{(2 \cdot 2)}=256$ and we aim to select $L_1'=L_2'=4$ input vectors for each user. Hence, the spatial coding rates for each user are equal to $r_{1_{\text{SC}}} = r_{2_{\text{SC}}}= \frac{\log_2(4)}{\log_2(16)}=0.5$.

\subsubsection{Optimization Problem} The first step of the spatial coding is performed to get a LUT of size $N \times 256$.
\subsubsection{Subset Selection}
The input vectors $\vect s_1$ and $\vect s_2$ for the two users belong to the following alphabet 
$$\mathcal{S}_1=\mathcal{S}_2 = D_G^{-1}\left\lbrace  \begin{bmatrix} 0\\0 \end{bmatrix}, \begin{bmatrix} 0\\1 \end{bmatrix}, \begin{bmatrix} 0\\2 \end{bmatrix}, \begin{bmatrix} 0\\3 \end{bmatrix}, \begin{bmatrix} 1\\0 \end{bmatrix}, ...,\begin{bmatrix} 3\\2 \end{bmatrix},\begin{bmatrix} 3\\3 \end{bmatrix}\right\rbrace ,$$
where 0, 1, 2 and 3 designate the four different QPSK symbols with Gray coding. We consider Table \ref{table:ss_example}, where the cost functions $\Phi$ for all possible combinations of the input vectors are stored. According to Algorithm \ref{algorithm:ss_mu}, we start with user 1 to select the desired vectors. First, we average the cost functions for each input vector $\vect s_1$ among all possible vectors $\vect s_2$, i.e. we take the average among the rows for each column. Second, we choose the 4 columns, i.e. input vectors $\vect s_1$, that have the best average cost functions ${\rm E}_{\vect{s}_2} \lbrace\Phi(\matt H,  \vect x, \vect { s})\rbrace$. Third, we average among the selected columns to get ${\rm E}_{\vect{s}_1} \lbrace\Phi(\matt H,  \vect x, \vect { s})\rbrace$. The 4 best average cost functions give us the input vectors $\vect s_2$ to select. Let us assume that we get the following subsets sorted in ascending order of the decimal value defined in (\ref{eq:decimal_value})
\begin{align*}
\mathcal{S}_1' &= D_G^{-1}\left\lbrace  \begin{bmatrix} 1\\0 \end{bmatrix}, \begin{bmatrix} 3\\1 \end{bmatrix},\begin{bmatrix} 0\\2 \end{bmatrix},  \begin{bmatrix} 2\\3 \end{bmatrix}\right\rbrace  \\
\mathcal{S}_2' &=  D_G^{-1}\left\lbrace \begin{bmatrix} 0\\0 \end{bmatrix},\begin{bmatrix} 1\\1 \end{bmatrix}, \begin{bmatrix} 2\\2 \end{bmatrix}, \begin{bmatrix} 3\\3 \end{bmatrix} \right\rbrace.
\end{align*}
In total we get $L' = L_1' \cdot L_2'=16$ possible input vectors $\vect s$. The corresponding optimal transmit vectors $\vect x$ are selected from the LUT to get the sub-LUT of size $N \times 16$.
\subsubsection{Coding and Mapping:} For coding the selected input vectors $\vect s_1 \in \mathcal{S}_1'$ and $\vect s_2 \in \mathcal{S}_2'$ we need only $\log_2(L_1')=\log_2(L_2')=2$ bits and we get the following encoding scheme for user 1
$$ 00 \longrightarrow \begin{bmatrix} 1\\0 \end{bmatrix}; \\
   01 \longrightarrow  \begin{bmatrix} 3\\1 \end{bmatrix}; \\
   10 \longrightarrow \begin{bmatrix} 0\\2 \end{bmatrix}; \\
   11 \longrightarrow  \begin{bmatrix} 2\\3 \end{bmatrix},$$
and for user 2
$$ 00 \longrightarrow \begin{bmatrix} 0\\0 \end{bmatrix}; \\
   01 \longrightarrow  \begin{bmatrix} 1\\1 \end{bmatrix}; \\
   10 \longrightarrow \begin{bmatrix} 2\\2 \end{bmatrix}; \\
   11 \longrightarrow  \begin{bmatrix} 3\\3 \end{bmatrix}.$$
According to the sub-LUT every input vector $\vect s = \begin{bmatrix}
\vect s_1^{\T} & \vect s_2^{\T}\end{bmatrix}^{\T}$ is mapped to its corresponding transmit vector $\vect x$.
\begin{table}
\centering
\begin{tabular}{| c|  c |c|c| c|  }
\hline
$\Phi(\matt H, \vect x, \vect { s})$ & $\vect s^{(1)}_1$ & $\cdots$ &  $\vect s^{(16)}_1$&${\rm E}_{\vect{s}_1} \lbrace\Phi(\matt H,  \vect x, \vect { s})\rbrace$\\
\hline
$\vect s^{(1)}_2$ &  & & &\\ \hline

 $\vdots$ &  & & &\\  \hline
 $\vect s^{(16)}_2$ & & &&\\ \hline
${\rm E}_{\vect{s}_2} \lbrace\Phi(\matt H,  \vect x, \vect { s})\rbrace$ & &&&\\\hline
\end{tabular}
\caption{Subset selection example.}
\label{table:ss_example}
\end{table}

\section{Simulation Results}
\label{sec:simresults}

We make use of LDPC code of length 256. The choice of this code length is motivated by having small latency time. The number of iterations of the LDPC decoding is fixed to 20 iterations.
The bit streams $\vect b_m$, $m=1,~\cdots,~M$ have a length of $N_b=10^6$.
The performance metric is the coded BER as function of the transmit power $P_{\text{tx}}$. We study the effect of the spatial coding rate on the performance for different correlation factors between the receive antennas of each user $\rho$ while keeping the total rate for each user constant, i.e. $r_m = 3/8$ for $m=1,~\cdots,~M$. In the simulations we consider the case of $N=64$ transmit antennas and $M=2$ users with $K=2$ receive antennas.

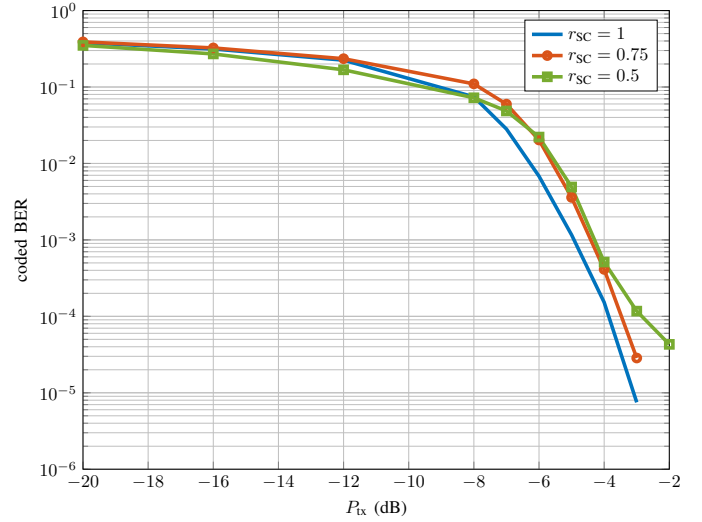
\begin{figure}
\centering
\resizebox{9cm}{!} {
% This file was created by matlab2tikz.
%
%The latest updates can be retrieved from
%  http://www.mathworks.com/matlabcentral/fileexchange/22022-matlab2tikz-matlab2tikz
%where you can also make suggestions and rate matlab2tikz.
%
\definecolor{mycolor1}{rgb}{0.00000,0.45098,0.74118}%
\definecolor{mycolor2}{rgb}{0.85098,0.32941,0.10196}%
\definecolor{mycolor3}{rgb}{0.47059,0.67059,0.18824}%
\begin{tikzpicture}

\begin{axis}[%
width=4.521in,
height=3.566in,
at={(0.758in,0.481in)},
scale only axis,
xmin=-20,
xmax=-2,
xlabel={$P_{\text{tx}}$ (dB)},
xmajorgrids,
ymode=log,
ymin=1e-06,
ymax=1,
yminorticks=true,
ylabel={coded BER},
ymajorgrids,
yminorgrids,
axis background/.style={fill=white},
legend style={legend cell align=left,align=left,draw=black}
]
\addplot [color=mycolor1,solid,line width=2.0pt]
  table[row sep=crcr]{%
-20	0.379457785298259\\
-16	0.31447012608807\\
-12	0.224367859543011\\
-8	0.0743607590885817\\
-7	0.0282053051395289\\
-6	0.00680093525985663\\
-5	0.0011625744047619\\
-4	0.00015250976062468\\
-3	7.50048003072197e-06\\
-2	0\\
-1	0\\
0	0\\
};
\addlegendentry{$r_{\text{SC}}=1$};

\addplot [color=mycolor2,solid,line width=2.0pt,mark=o,mark options={solid}]
  table[row sep=crcr]{%
-20	0.389345918138761\\
-16	0.324588773681516\\
-12	0.234639016897081\\
-8	0.109870031682028\\
-7	0.0596818196364567\\
-6	0.0202667970750128\\
-5	0.00359272993471582\\
-4	0.000411026305683564\\
-3	2.85018241167435e-05\\
-2	0\\
-1	0\\
0	0\\
};
\addlegendentry{$r_{\text{SC}}=0.75$};
\addplot [color=mycolor3,solid,line width=2.0pt,mark=square,mark options={solid}]
  table[row sep=crcr]{%
-20	0.350013900889657\\
-16	0.270849334357399\\
-12	0.167868243567588\\
-8	0.0722786258320532\\
-7	0.048578609030978\\
-6	0.0221419170826933\\
-5	0.00492481518817204\\
-4	0.000513032834101382\\
-3	0.000117007488479263\\
-2	4.30027521761393e-05\\
-1	0\\
0	0\\
};
\addlegendentry{$r_{\text{SC}}=0.5$};
\end{axis}
\end{tikzpicture}%
}
\caption{MU scenario with $N=64,~M=2,~K=2,~\rho=0.2$ and $r=3/8$.}
\label{fig:rho_0.2}
\end{figure}

\begin{figure}
\centering
\resizebox{9cm}{!} {
% This file was created by matlab2tikz.
%
%The latest updates can be retrieved from
%  http://www.mathworks.com/matlabcentral/fileexchange/22022-matlab2tikz-matlab2tikz
%where you can also make suggestions and rate matlab2tikz.
%
\definecolor{mycolor1}{rgb}{0.85098,0.32941,0.10196}%
\definecolor{mycolor2}{rgb}{0.47059,0.67059,0.18824}%
\begin{tikzpicture}

\begin{axis}[%
width=4.521in,
height=3.566in,
at={(0.758in,0.481in)},
scale only axis,
xmin=-11,
xmax=10,
xlabel={$P_{\text{tx}}$ (dB)},
xmajorgrids,
ymode=log,
ymin=1e-06,
ymax=1,
yminorticks=true,
ylabel={Coded BER},
ymajorgrids,
yminorgrids,
axis background/.style={fill=white},
legend style={legend cell align=left,align=left,draw=black}
]
\addplot [color=blue,solid,line width=2.0pt]
  table[row sep=crcr]{%
-11	0.240414386520737\\
-7	0.199028737839222\\
-3	0.166138132840502\\
1	0.095817132296467\\
2	0.0697569644457245\\
3	0.0439078100998464\\
4	0.0234490007360471\\
5	0.0103881648425499\\
6	0.00380824372759857\\
7	0.00113357254864311\\
8	0.000266517057091654\\
9	7.6504896313364e-05\\
};
\addlegendentry{$r_{\text{SC}}=1$};

\addplot [color=mycolor1,solid,line width=2.0pt,mark=o,mark options={solid}]
  table[row sep=crcr]{%
-11	0.179181467613927\\
-7	0.0868560587877624\\
-3	0.0353517625128008\\
1	0.0110217053891449\\
2	0.00705445148489503\\
3	0.00431377608166923\\
4	0.00245015681003584\\
5	0.00145259296594982\\
6	0.000768049155145929\\
7	0.000436527937788018\\
8	0.000183511744751664\\
9	3.55022721454173e-05\\
};
\addlegendentry{$r_{\text{SC}}=0.75$};
\addplot [color=mycolor2,solid,line width=2.0pt,mark=square,mark options={solid}]
  table[row sep=crcr]{%
-19	0.258083017313108\\
-15	0.153099298355095\\
-11	0.0579927115335381\\
-7	0.000152009728622632\\
-6	3.00019201228879e-06\\
-5	0\\
-4	0\\
-3	0\\
-2	0\\
-1	0\\
0	0\\
1	0\\
};
\addlegendentry{$r_{\text{SC}}=0.5$};
\end{axis}
\end{tikzpicture}%
}
\caption{MU scenario with $N=64,~M=2,~K=2,~\rho=0.8$ and $r=3/8$.}
\label{fig:rho_0.8}
\end{figure}
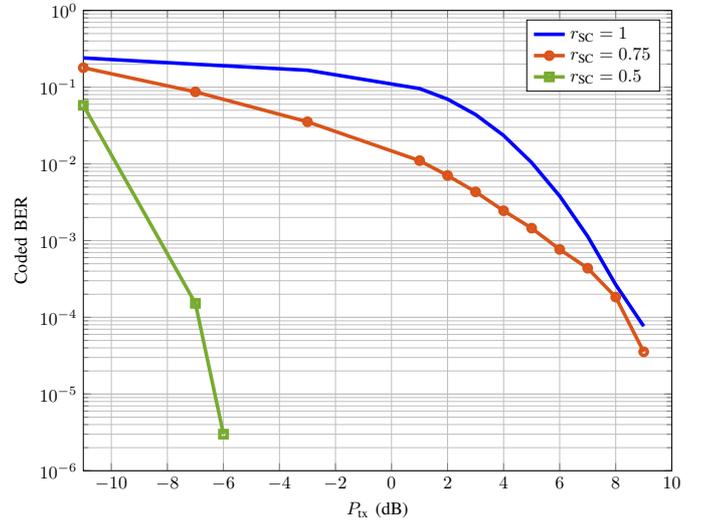

As shown in Fig. \ref{fig:rho_0.2} the spatial coding does not lead to any performance improvement when $\rho=0.2$. However, when the correlation factor is large $\rho=0.8$, as depicted in Fig. \ref{fig:rho_0.8}, the spatial coding with $r_{\text{SC}}=0.5$ achieves a gain of more than 12dB. When the receive antennas at each user are highly correlated, some vectors from $\mathcal{S}$ cannot be detected. Therefore, it is beneficial to get rid of those vectors, that are badly transmitted through the channel and hence badly detected at the receiver.

\section{Conclusion}
We presented a spatial coding technique, that can significantly improve the performance in terms of coded BER, up to 12dB. This gain is obtained for highly correlated channels, where the conventional coding techniques might reach their limits.

Other methods for the subset selection and the coding of the selected vectors can be considered to further improve the performance. Furthermore,
this spatial coding can be applied combined with other precoding design criteria like symbol-wise minimum square error (MSE).
\label{sec:conclusion}
\bibliographystyle{IEEEtran}
\bibliography{IEEEabrv,refs}

% Generated by IEEEtran.bst, version: 1.13 (2008/09/30)
\begin{thebibliography}{1}
\providecommand{\url}[1]{#1}
\csname url@samestyle\endcsname
\providecommand{\newblock}{\relax}
\providecommand{\bibinfo}[2]{#2}
\providecommand{\BIBentrySTDinterwordspacing}{\spaceskip=0pt\relax}
\providecommand{\BIBentryALTinterwordstretchfactor}{4}
\providecommand{\BIBentryALTinterwordspacing}{\spaceskip=\fontdimen2\font plus
\BIBentryALTinterwordstretchfactor\fontdimen3\font minus
  \fontdimen4\font\relax}
\providecommand{\BIBforeignlanguage}[2]{{%
\expandafter\ifx\csname l@#1\endcsname\relax
\typeout{** WARNING: IEEEtran.bst: No hyphenation pattern has been}%
\typeout{** loaded for the language `#1'. Using the pattern for}%
\typeout{** the default language instead.}%
\else
\language=\csname l@#1\endcsname
\fi
#2}}
\providecommand{\BIBdecl}{\relax}
\BIBdecl

\bibitem{5G}
R.~Baldemair, E.~Dahlman, G.~Fodor, G.~Mildh, S.~Parkvall, Y.~Selen,
  H.~Tullberg, and K.~Balachandran, ``Evolving wireless communications:
  Addressing the challenges and expectations of the future,'' \emph{IEEE
  Vehicular Technology Magazine}, vol.~8, no.~1, pp. 24--30, March 2013.

\bibitem{LarssonMarzetta2014}
E.~G. Larsson, O.~Edfors, F.~Tufvesson, and T.~L. Marzetta, ``Massive mimo for
  next generation wireless systems,'' \emph{IEEE Communications Magazine},
  vol.~52, no.~2, pp. 186--195, February 2014.

\bibitem{mmWave5G}
Z.~Pi and F.~Khan, ``An introduction to millimeter-wave mobile broadband
  systems,'' \emph{IEEE Communications Magazine}, vol.~49, no.~6, pp. 101--107,
  June 2011.

\bibitem{SM2001}
Y.~A. Chau and S.-H. Yu, ``Space modulation on wireless fading channels,'' in
  \emph{Vehicular Technology Conference, 2001. VTC 2001 Fall. IEEE VTS 54th},
  vol.~3, 2001, pp. 1668--1671 vol.3.

\bibitem{SMSiemens2012}
H.~Haas, E.~Costa, and E.~Schulz, ``Increasing spectral efficiency by data
  multiplexing using antenna arrays,'' in \emph{Personal, Indoor and Mobile
  Radio Communications, 2002. The 13th IEEE International Symposium on},
  vol.~2, Sept 2002, pp. 610--613 vol.2.

\bibitem{JeddaSAM2016}
H.~Jedda, A.~Mezghani, and J.~A. Nossek, ``Minimum ber precoding in 1-bit
  massive mimo systems,'' in \emph{Sensor Array and Multichannel Signal
  Processing Workshop (SAM), 2016 IEEE 9th}, July 2016.

\end{thebibliography}
\end{document}